\documentclass{webofc}
\usepackage[varg]{txfonts}   % Web of Conferences font
\usepackage{caption}
\usepackage{subcaption}

\begin{document}
\title{A DNN for CMS track classification and selection}

\author{\firstname{Leonardo} \lastname{Giannini}\inst{1}\fnsep\thanks{\email{leonardo.giannini@cern.ch}}  on behalf of the CMS Collaboration}

\institute{University of California, San Diego}

\abstract{
The upgrade of the track classification and  selection step of the CMS tracking to a Deep Neural Network is presented. 
The CMS tracking follows an iterative approach: tracks are reconstructed in multiple passes starting from the ones that are easiest to find and moving to the ones with more complex characteristics (lower transverse momentum, high displacement). 
The track classification comes into play at the end of each iteration.
A classifier using a multivariate analysis is applied after each iteration and several selection criteria are defined. If a track meets the high purity requirement, its hits are removed from the hit collection, thus simplifying the later iterations, and making the track classification an integral part of the reconstruction process. Tracks passing loose selections are also saved for physics analysis usage.
The CMS experiment improved the track classification starting from a parametric selection used in Run 1, moving to a Boosted Decision Tree in Run 2, and finally to a  Deep Neural Network in Run 3. An overview of the Deep Neural Network training and current performance is shown.
}
\maketitle 
\section{Introduction}\label{se1}

An iterative approach is employed for the CMS tracking \cite{CMS:2014pgm}: the track reconstruction is run several times, starting from easier tracks, i.e. non-displaced tracks and with relatively high transverse momentum, and progressively moving to the more complex ones, as the hits from high quality tracks are masked after each iteration for the later ones. This iterative approach is aimed to ensure the best possible efficiency while keeping low fake rates. The four main steps repeated in each iteration can be summarized as follows:
\begin{itemize}
\item Track seeding - Seeding defines an initial tracklet with its parameters and uncertainties. The seeding needs at least three 3D hits (or two with an additional constraint on the origin of the track). Pixel hits are preferred for seeding because of the better resolution and the lower occupancy of the pixel detector compared to the strips. They are also less affected by interaction in the tracker material. Strip "stereo" hits are also used, but not in the initial tracking iterations.
Different types of seeds are used by each iteration: the initial iteration uses seeds with four pixel hits, relatively high transverse momentum, then three pixel hits and lower transverse momentum seeds are used. Finally seeds with some or all the hits in the strip detectors are used to build full tracks.
\item Pattern recognition or Track finding/building - Compatible hits are associated 
to the seed, based on Kalman filter techniques. The seed trajectories are propagated searching for compatible hits in the outer layers. The seeds are extrapolated layer by layer through the tracker and at each layer the track parameters are updated. The material crossed
by the track and the uncertainties of the hits are taken into account. Ghost hits, corresponding to layers without measured charge deposits, but whose material
needs to be considered, are also added to account for possible tracker inefficiencies.
Once the track is completed another search is performed backwards starting from the
outermost hit. This step is performed because the pattern recognition is more efficient than the seeding, as it handles correctly the possibility that the silicon modules overlap, and it
can recover pixel hits which were not used at seeding time. 
In case multiple compatible hits are found when extrapolating the trajectory to a single layer, the algorithm will create one trajectory candidate for each hit and those are
propagated independently. Eventually, only one track is retained based on the quality
and the total number of hits.
\item Track fitting - Once the tracks are built the trajectory is refitted by combining inside-out and outside-in results at each hit in a smoothing procedure. The smoothed trajectory predicted at each hit allows to optimally select outlier hits. After each outlier hit is removed the procedure is repeated until no outlier hits are left.
\item Track selection -  At the end of each iteration the fitted tracks are classified and divided into several categories (high purity, loose, ...). The track selection is based on several track parameters and was initially implemented as a parametric selection in Run 1 \cite{CMS:2014pgm}, moving to a Boosted Decision Tree (BDT) in Run 2. A Deep Neural Network (DNN), presented in this proceeding, replaces the Run 2 BDT \cite{CMS-DP-2023-009}.
The track selection is an integral part of the iterative tracking, as hits coming from high purity tracks are removed for the subsequent iterations, thus reducing the combinatorics of the pattern recognition.
\end{itemize}

After all the iterations are completed, the tracks are merged into a single collection. Tracks passing loose requirements are selected for this single collection. Tighter requirements can be applied to select higher quality tracks, e.g. the particle flow reconstruction \cite{CMS:2017yfk} uses high purity tracks.\\

The track finding step underwent a significant change between the end of the LHC Run 2 an the beginning of the Run 3. The standard tracking algorithm used by CMS in Run 1 and Run 2 is the so-called Combinatorial Kalman Filter (CKF) \cite{CMS:2014pgm}. For some of the tracking iterations, the parallelized and vectorized version of the Kalman Filter (mkFit) was introduced \cite{Lantz:2020yqe,CMS-DP-2022-018}. 
In the current default tracking the InitialStepPreSplitting, Initial, HighPtTriplet, DetachedQuad, DetachedTriplet iterations use mkFit. The rest uses CKF as in Run 2.
At the same time the track selection was updated from a BDT to a DNN. The DNN was developed initially for pure CKF reconstruction, but later for both mkFit and CKF tracks, which can have slight differences.
%, most notably in the total number of hits. 
These differences required training the DNN twice using two different sets of input tracks, while keeping all the training settings unchanged. 
In the current default tracking a DNN trained on mkFit tracks is used for the mkFit iterations, while a DNN trained on the CKF tracks is used for the CKF iterations.\\

In the following sections the DNN training strategy is described in detail and its performance is compared to the BDT performance with some caveats. The BDT was trained on CKF only, but the conditions were different at the beginning of Run 2 (lower pileup and center of mass energy). The DNN performance is shown in the current default tracking (mkFit + CKF Run 3) and compared to the result of the Run 2 BDT on the same tracks. The same comparison could be more coherent with a BDT retraining on the same set of tracks, but that additional task wasn't worth pursuing for pure comparison reasons.

\section{The Track selection Deep Neural Network}

%The track selection DNN is introduced for Run 3 tracking \cite{CMS-DP-2023-009}. Previously a BDT was employed in Run 2 and a parametric selection in Run 1 \cite{CMS:2014pgm}.\\

The track selection DNN is trained using reconstructed tracks produced in simulations. High level input features, which characterize the entire track, are used in training. The inputs are both discriminating features and context features like the p$_\text{T}$, $\eta$ and $\phi$ of the tracks. The full set of the features fed to the DNN is listed below:
\begin{itemize}
\item[-] the track p$_\text{T}$
, $\eta$, $\phi$, and their respective uncertainties $\delta$p$_\text{T}$
, $\delta\eta$, $\delta\phi$
\item[-] p$_\text{X}$
, p$_\text{Y}$
, p$_\text{Z}$
, p$_\text{T}$
for the innermost and outermost state of the track
\item[-] the transverse and longitudinal impact parameters, d$_0$, d$_\text{Z}$, computed both from the beamspot and from the closest primary vertex, and
their respective uncertainties $\delta$d$_0$, $\delta$d$_\text{Z}$
\item[-] the track $\chi^{2}$ and number of degrees of freedom
\item[-] number of hits in the pixel detector, number of hits in the strip detector
\item[-] number of missing hits inside the innermost hit and outside the 
outermost hit
\item[-] number of inactive layers crossed inside the innermost hit and outside
the outermost hit
\item[-] number of layers without hits overall
\item[-] the iteration index
\end{itemize}
The target of the training is a true or false label for each track reconstructed in simulation. A track must have more than 75\% of its hits matched to a single simulated track to be labeled as true \cite{CMS:2014pgm}. \\

\begin{figure}[h!]
\centering
\includegraphics[width=0.6\textwidth]{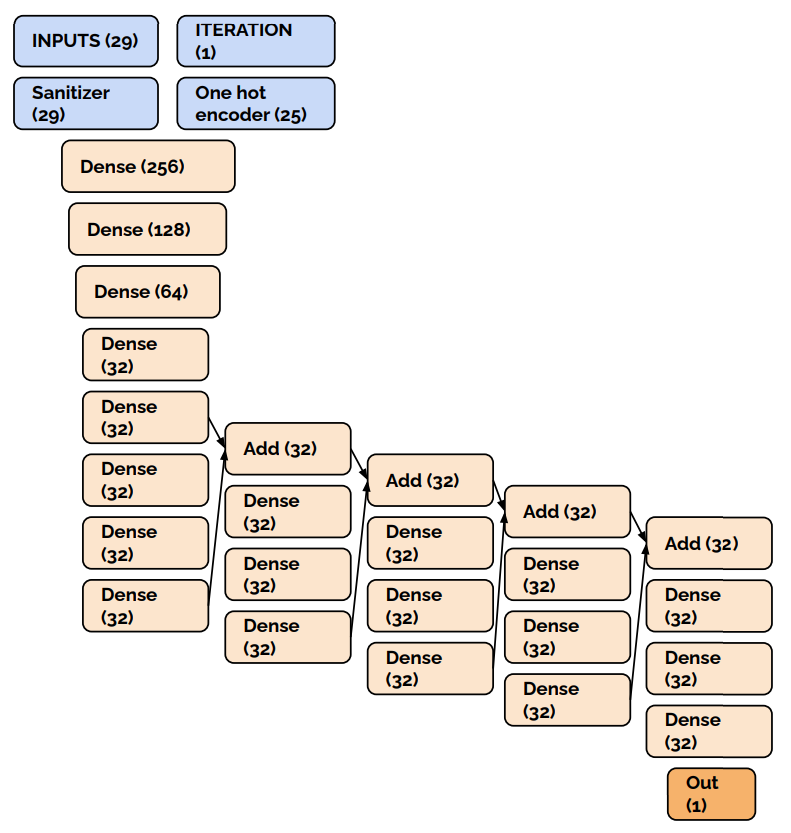}
\caption{Schematic representation of the DNN architecture}\label{figs}       
\end{figure}

The neural network chosen for the training is a simple feed-forward network, whose architecture is shown in figure \ref{figs}. The “sanitizer” layer shown in the scheme applies logarithmic and/or absolute value transformations to some of the input features, while the “one hot encoder” layer converts the integer iteration index into a boolean vector with a true value corresponding to the iteration index. The numbers in the dense blocks indicate the number of nodes used in the DNN layer. The network has 5 feed-forward layers followed by 5 “skip-connection” blocks. The input and the output of these blocks are summed and passed as input to the following “skip-connection” block. 
The activation functions in the hidden layers are all ELUs \cite{clevert2016fast}, while a sigmoid is used in the output layer. The loss function is the binary cross-entropy. The training was performed using the \textsc{Keras} software package \cite{chollet2015keras}  with \textsc{TensorFlow} backend \cite{abadi2016tensorflow}.\\

The training was performed on tracks, including those from pileup vertices, from several simulated samples generated at a center-of-mass energy of 14 TeV. All of them have a pileup distribution sampled from minimum bias events with Poisson mean distributed between 20 and 70.
The dataset specifically includes:
\begin{itemize}
\item QCD multijet production - generated with a flat hard-scattering p$_\text{T}$ from 15 GeV to 7 TeV
\item  t$\bar{\text{t}}$ production
\item  Drell-Yan production with Z decaying into electrons
\item  Stop-antistop ($\tilde{\text{t}}\bar{\tilde{\text{t}}}$) production in RPV SUSY, with stop masses of 1 TeV and 1.8 TeV and stop decay lengths of 10 or 100 cm. These samples are used to increase the amount of displaced tracks
\end{itemize}

No track selection is applied to tracks used in training: all the tracks are labeled as “high purity” and the hit masking for the later iterations uses all the tracks from the previous iterations. This is a reasonable approximation and it allows to train for all the tracking iteration in a single step, while being agnostic to the previous version of the discriminator. The training sample size used in this round of training is 1.3B tracks. Five training epochs are run over the sample. A batch size of 512 is used for the training, with the Adam optimizer \cite{kingma2017adam} being employed.\\

After the training is performed the working points are chosen iteration by
iteration in a validation sample similar to the training one. The efficiency is set to approximately match the Run 2 BDT efficiency, with a possible fake rate reduction. The choice of the working point is validated in standard tracking with the hit masking applied.

\section{Tracking Performance}

The tracking performance is evaluated in terms of tracking efficiency, fake rate and duplicate rate evaluated in simulated samples, by associating reconstructed tracks and simulated tracks. A reconstructed track is considered associated with a simulated particle if more than 75\% of its hits originate from this
simulated particle \cite{CMS:2014pgm}. If this is not the case, the reconstructed track is considered as a random combination of hits and marked as a fake track.\\

Simulated tracks coming from the signal (hard scattering) vertex are used in the efficiency computation. The tracking efficiency is defined as the fraction of simulated tracks associated with at least one reconstructed track.
All reconstructed tracks coming from any vertex (including pileup vertices) are used in the fake rate and duplicate rate computation. The tracking fake rate is defined as the fraction of misidentified reconstructed tracks; the tracking duplicate rate is defined as the fraction of reconstructed tracks associated multiple times to the same simulated track.\\

The performance has been measured in both a t$\bar{\text{t}}$ sample and in a sample with $\tilde{\text{t}}\bar{\tilde{\text{t}}}$ (stop-antistop) production in RPV SUSY, similar to the ones used in training, where the stops have a significant decay length and produce displaced tracks, i.e: pp$\rightarrow\tilde{\text{t}}\bar{\tilde{\text{t}}}$  with , $\tilde{\text{t}}\rightarrow\ell\text{b}$,  m($\tilde{\text{t}}$) = 800 GeV, c$\tau$($\tilde{\text{t}}$) = 50 cm, with superimposed pileup events. 
The number of pileup events is sampled from minimum bias events with Poisson mean flatly distributed from from 55 to 75. 
The detector conditions match the most recent Run 3 simulation, as of February 2023. The efficiency, fake rate, duplicate rate are shown
as a function of p$_\text{T}$, $\eta$, pileup for the t$\bar{\text{t}}$ sample, while the same quantities are shown as a function of the track displacement for the $\tilde{\text{t}}\bar{\tilde{\text{t}}}$ sample.
The physics results are shown after applying the high purity BDT or DNN selection to each iteration and after merging all the tracks from
the iterations into one collection. The BDT results are shown in this comparison with the caveat explained in section \ref{se1}.\\

%The tracking efficiency is consistent or slightly better, when comparing the DNN to the Run 2 BDT\\

%The tracking fake
%rate is overall
%lower. Most
%notably in the high
%and low pT
%range,
%in the barrel and
%encap ($|\eta|$<1 or
%$|\eta|$>2 ) and at
%higher PU values
%\\

%The duplicate rate
%is about 20\% higher
%(see ref. [2]), due to
%the merging of tracks
%selected by different
%DNNs trained on
%mkFit or legacy CKF
%reconstruction

\begin{figure}[h!]
\centering
    \begin{subfigure}[b]{0.3\textwidth}
    \centering
    \includegraphics[width=\textwidth]{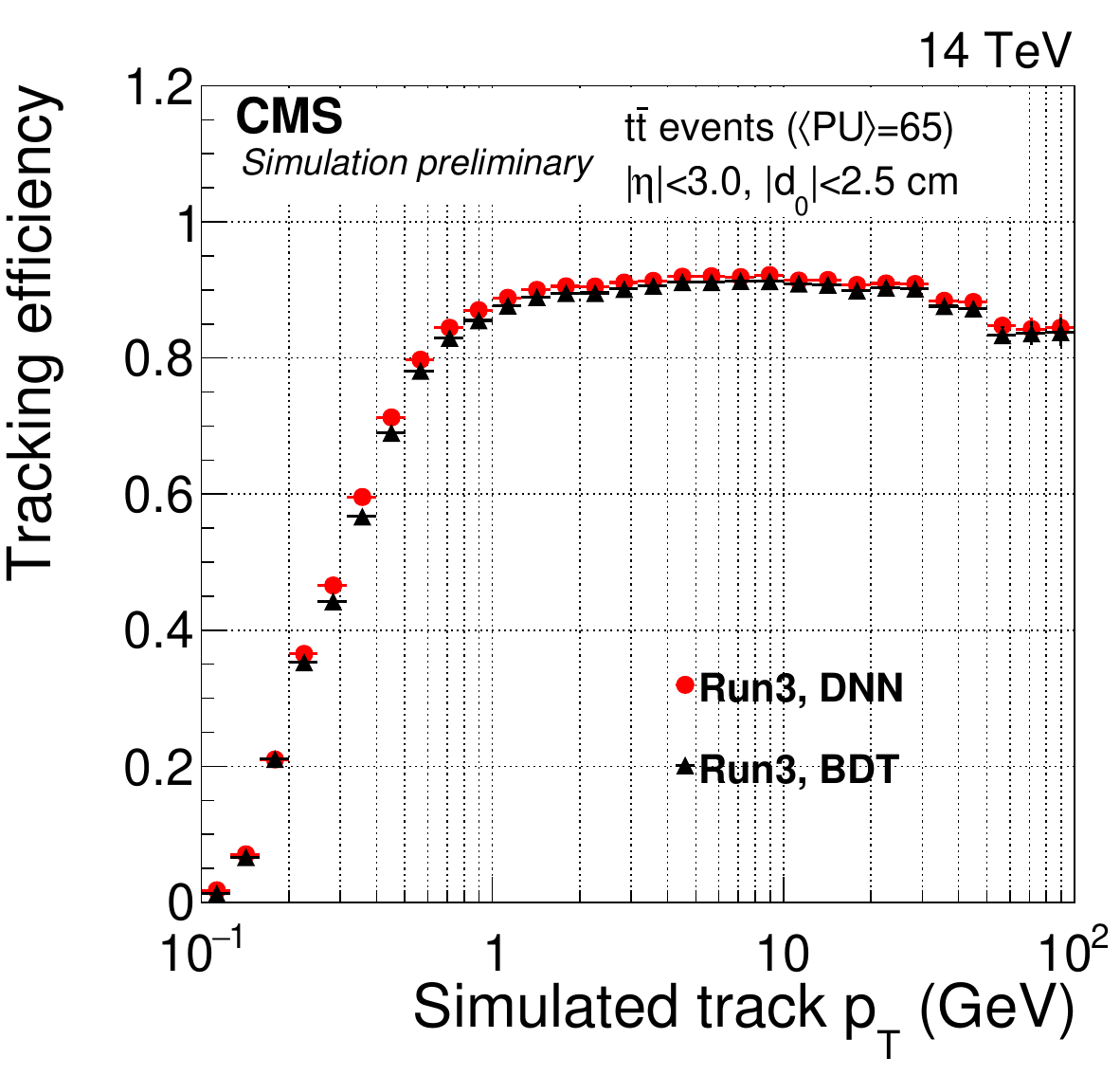}
    \caption{\label{fig:image1}}
    \end{subfigure}
\quad
    \begin{subfigure}[b]{0.3\textwidth}
    \centering
    \includegraphics[width=\textwidth]{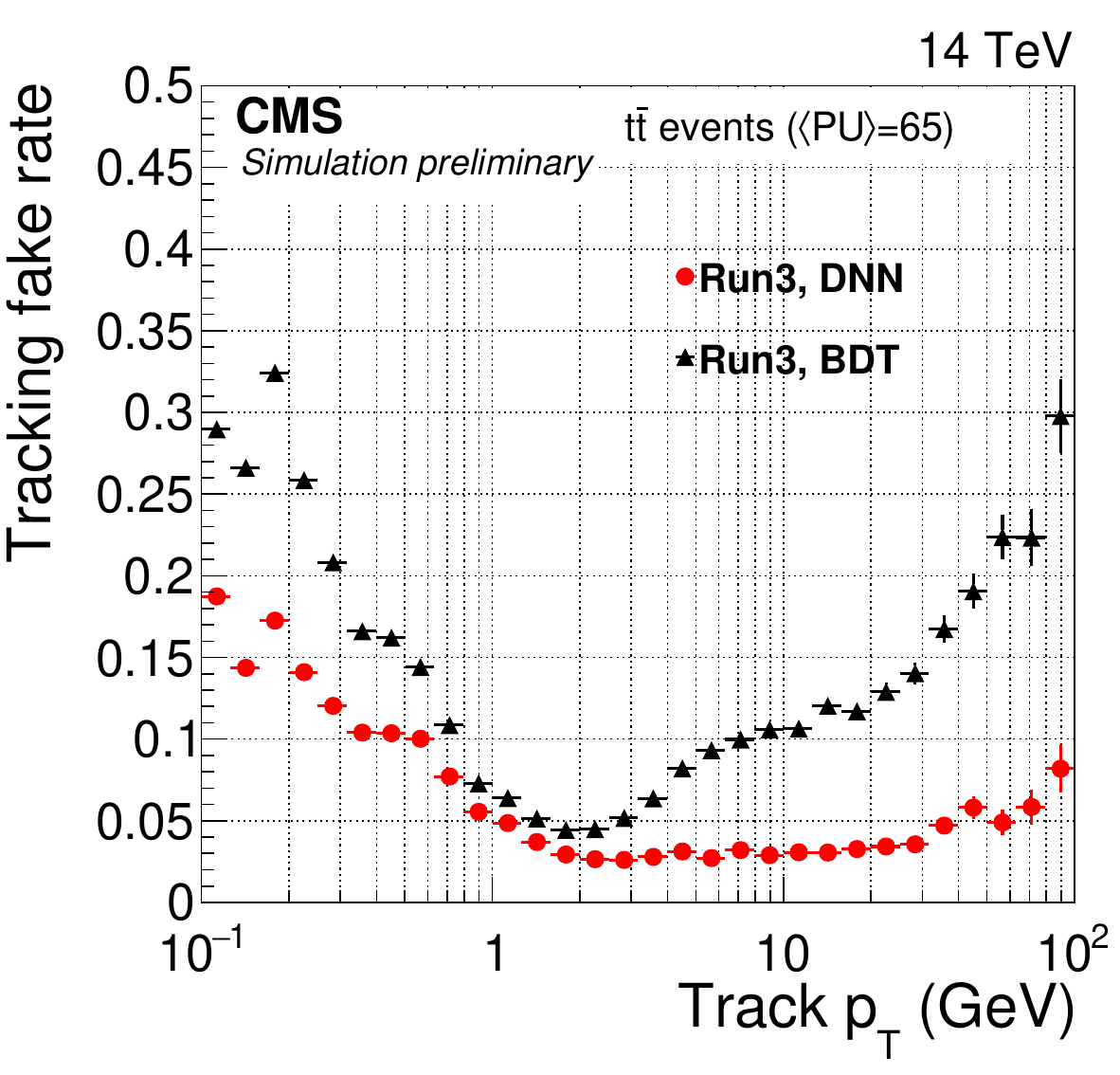}
    \caption{\label{fig:image2}}
    \end{subfigure}
\quad
    \begin{subfigure}[b]{0.3\textwidth}
    \centering
    \includegraphics[width=\textwidth]{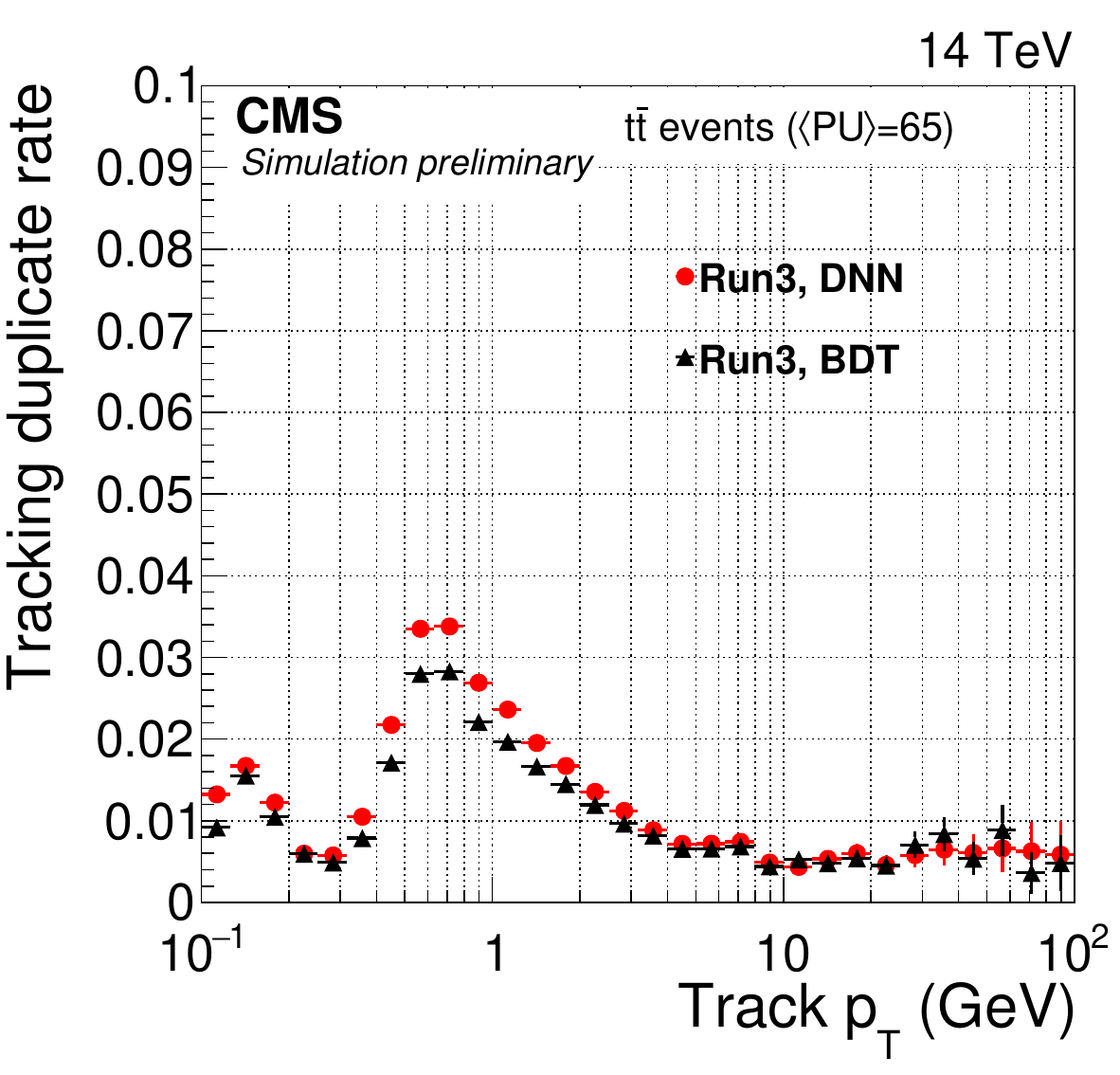}
    \caption{\label{fig:image3}}
    \end{subfigure}
\caption{(\subref{fig:image1}) Tracking efficiency as a function of
the simulated track p$_\text{T}$ for simulated tracks with $|\eta|$ < 3.0 and $|\text{d}_0|$ < 2.5 cm, (\subref{fig:image2}) tracking fake rate as a function of
the reconstructed track p$_\text{T}$, (\subref{fig:image3}) tracking duplicate rate as a function of the reconstructed track p$_\text{T}$.}\label{fig:pt}
\end{figure}

The efficiency, fake rate, duplicate rate versus p$_\text{T}$ are shown in figure \ref{fig:pt}. The tracking efficiency when the DNN is used is consistent with or slightly higher than the one obtained using the BDT across the entire  p$_\text{T}$
range. The efficiency improves the most at low  p$_\text{T}$, up to 5\%.  The tracking fake rate when the DNN is used is notably lower than the one obtained using the
BDT, especially for very low and very high p$_\text{T}$ values. Overall the fake rate is reduced by about 40\%. The tracking duplicate rate is slightly increased
when the DNN is used instead of the BDT. The overall increase is around 20\%. A slightly higher duplicate rate is expected due to merging of mkFit and CKF tracks selected by
different DNNs.\\

\begin{figure}[h!]
\centering
    \begin{subfigure}[b]{0.3\textwidth}
    \centering
    \includegraphics[width=\textwidth]{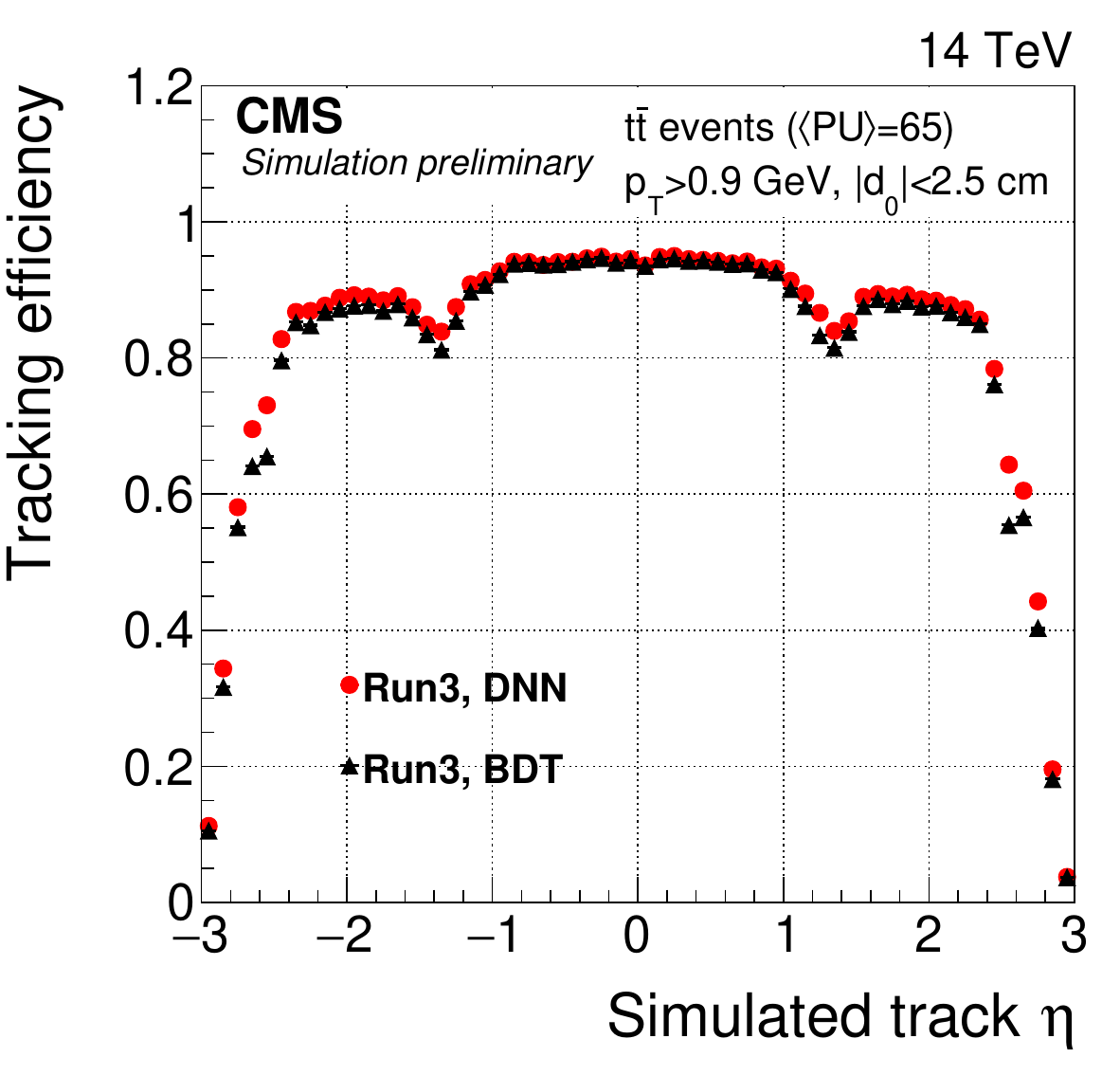}
    \caption{\label{fig:image1}}
    \end{subfigure}
\quad
    \begin{subfigure}[b]{0.3\textwidth}
    \centering
    \includegraphics[width=\textwidth]{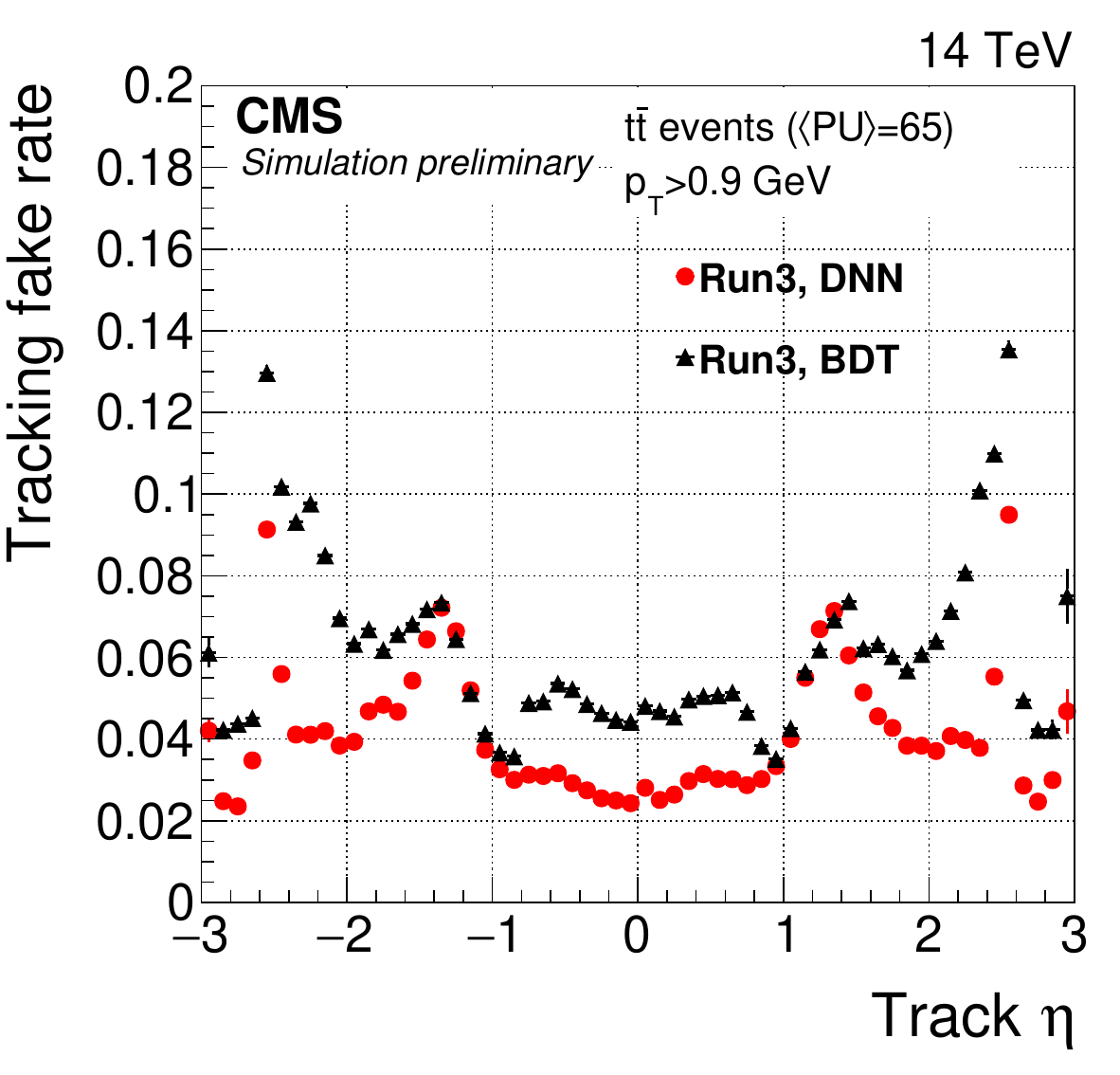}
    \caption{\label{fig:image2}}
    \end{subfigure}
\quad
    \begin{subfigure}[b]{0.3\textwidth}
    \centering
    \includegraphics[width=\textwidth]{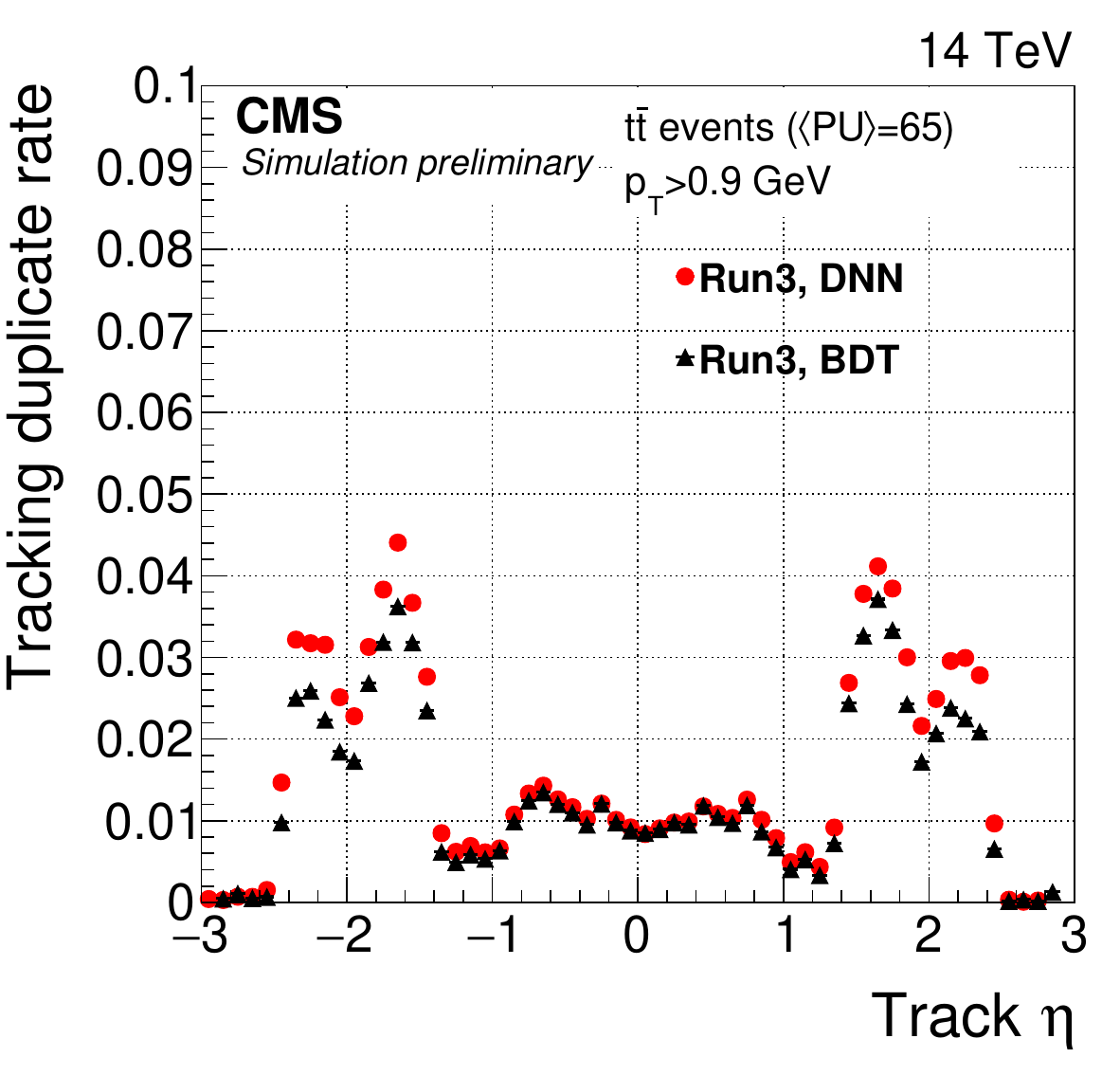}
    \caption{\label{fig:image3}}
    \end{subfigure}
\caption{(\subref{fig:image1}) Tracking efficiency as a function of
the simulated track $\eta$ for simulated tracks with p$_\text{T}$ > 0.9 GeV and $|\text{d}_0|$ < 2.5 cm, (\subref{fig:image2}) tracking fake rate as a function of the reconstructed track $\eta$ for
tracks with p$_\text{T}$ > 0.9 GeV, (\subref{fig:image3}) tracking duplicate rate as a function of the reconstructed track $\eta$ for
tracks with p$_\text{T}$ > 0.9 GeV.}\label{fig:eta}
\end{figure}

The efficiency, fake rate, duplicate rate versus $\eta$ are shown in figure \ref{fig:eta}.
The tracking efficiency when the DNN is used is
consistent with or slightly higher than the one
obtained using the BDT in all the $\eta$ regions. The
improvement in efficiency is at most 2\%. 
The tracking fake rate when the DNN is used is
lower than or consistent with the one obtained using
the BDT. The largest fake rate reductions are in
the tracker endcaps ($|\eta|$ > 2) and in the barrel
($|\eta|$ > 1). The tracking duplicate rate when the DNN is used
is higher than or consistent with the one obtained
using the BDT. In particular the increase is visible
in the endcap and transition regions ($|\eta|$ > 1). The
increase is up to 20\%.\\

\begin{figure}[h!]
\centering
    \begin{subfigure}[b]{0.3\textwidth}
    \centering
    \includegraphics[width=\textwidth]{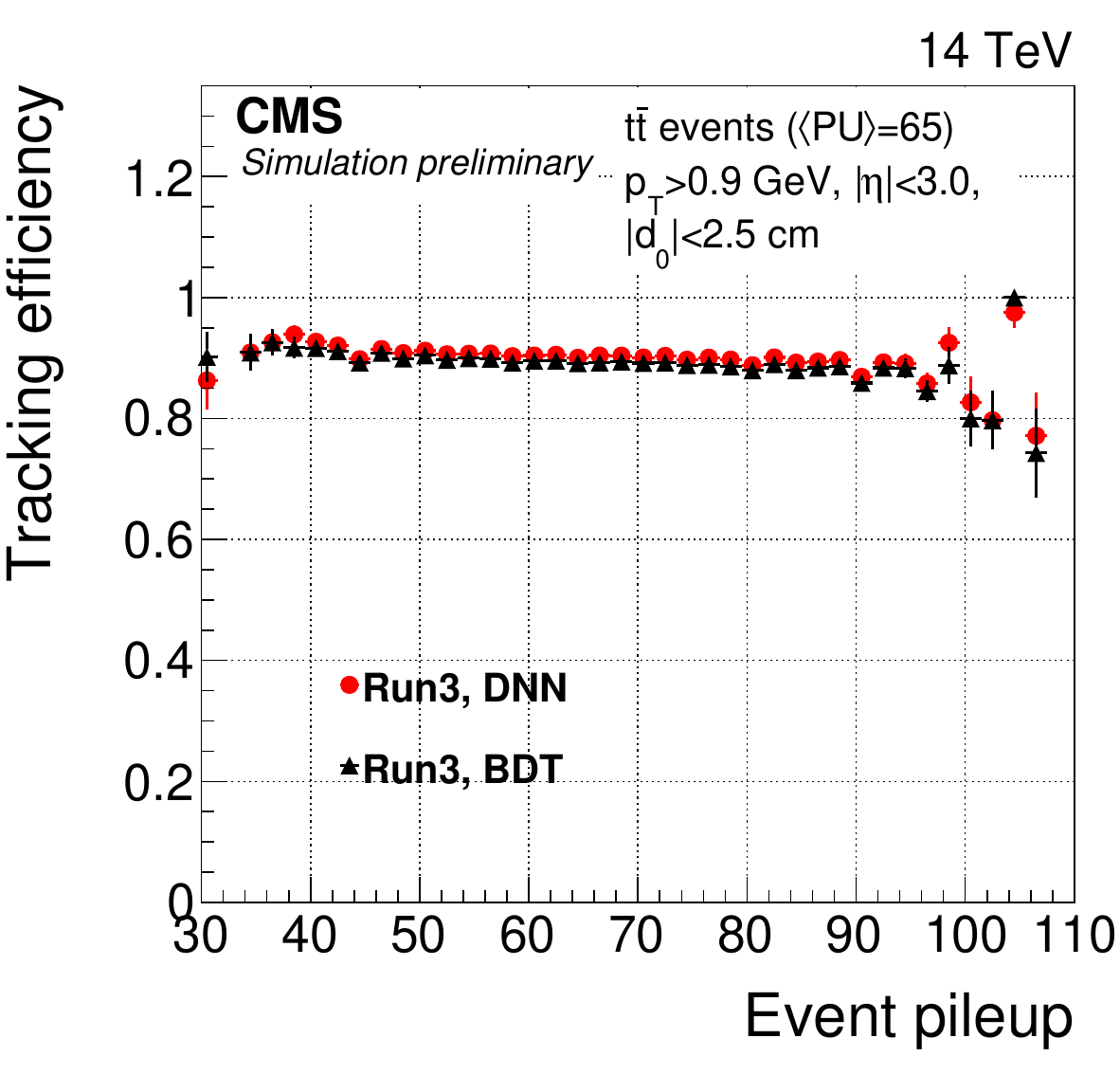}
    \caption{\label{fig:image1}}
    \end{subfigure}
\quad
    \begin{subfigure}[b]{0.3\textwidth}
    \centering
    \includegraphics[width=\textwidth]{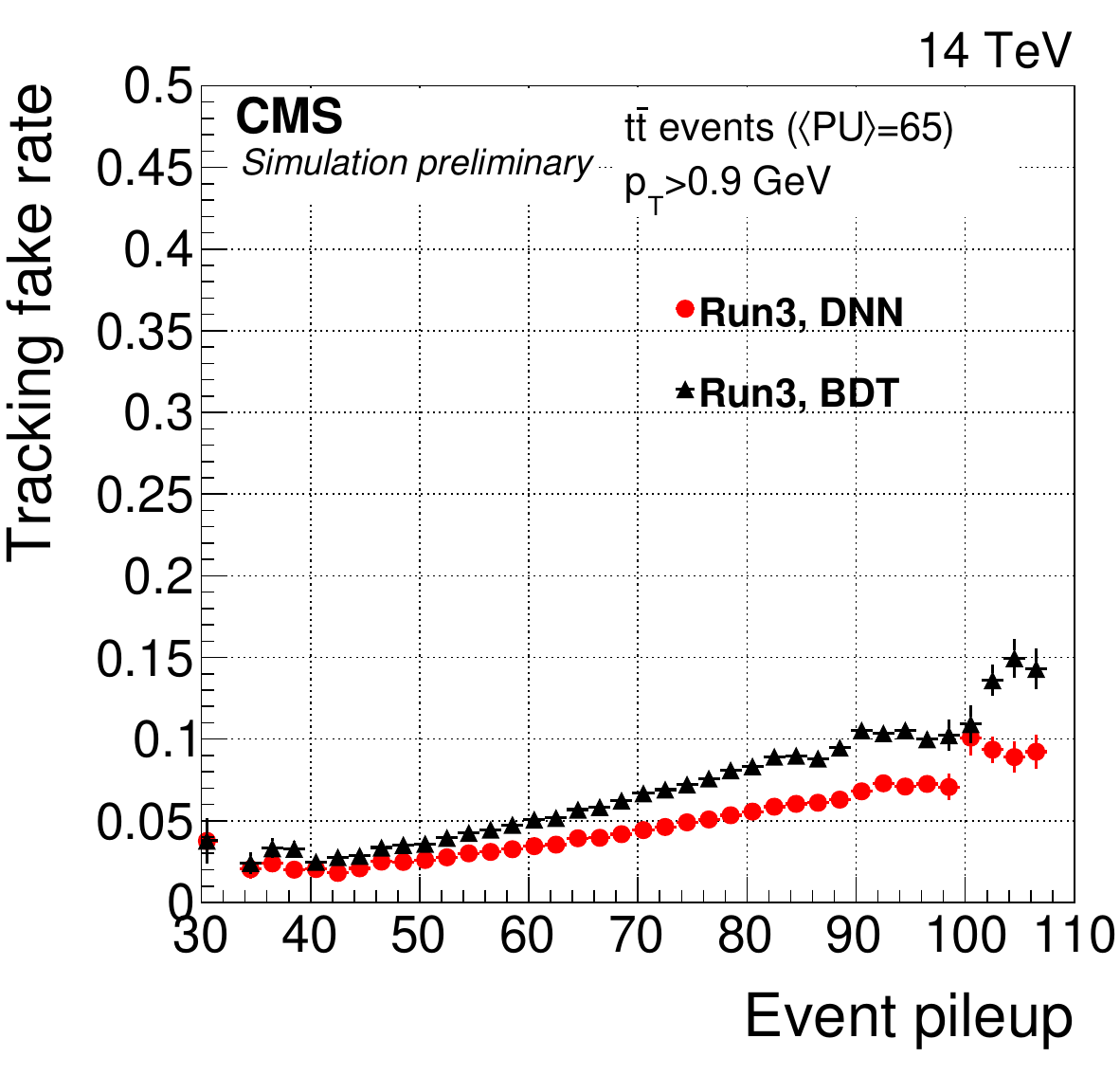}
    \caption{\label{fig:image2}}
    \end{subfigure}
\quad
    \begin{subfigure}[b]{0.3\textwidth}
    \centering
    \includegraphics[width=\textwidth]{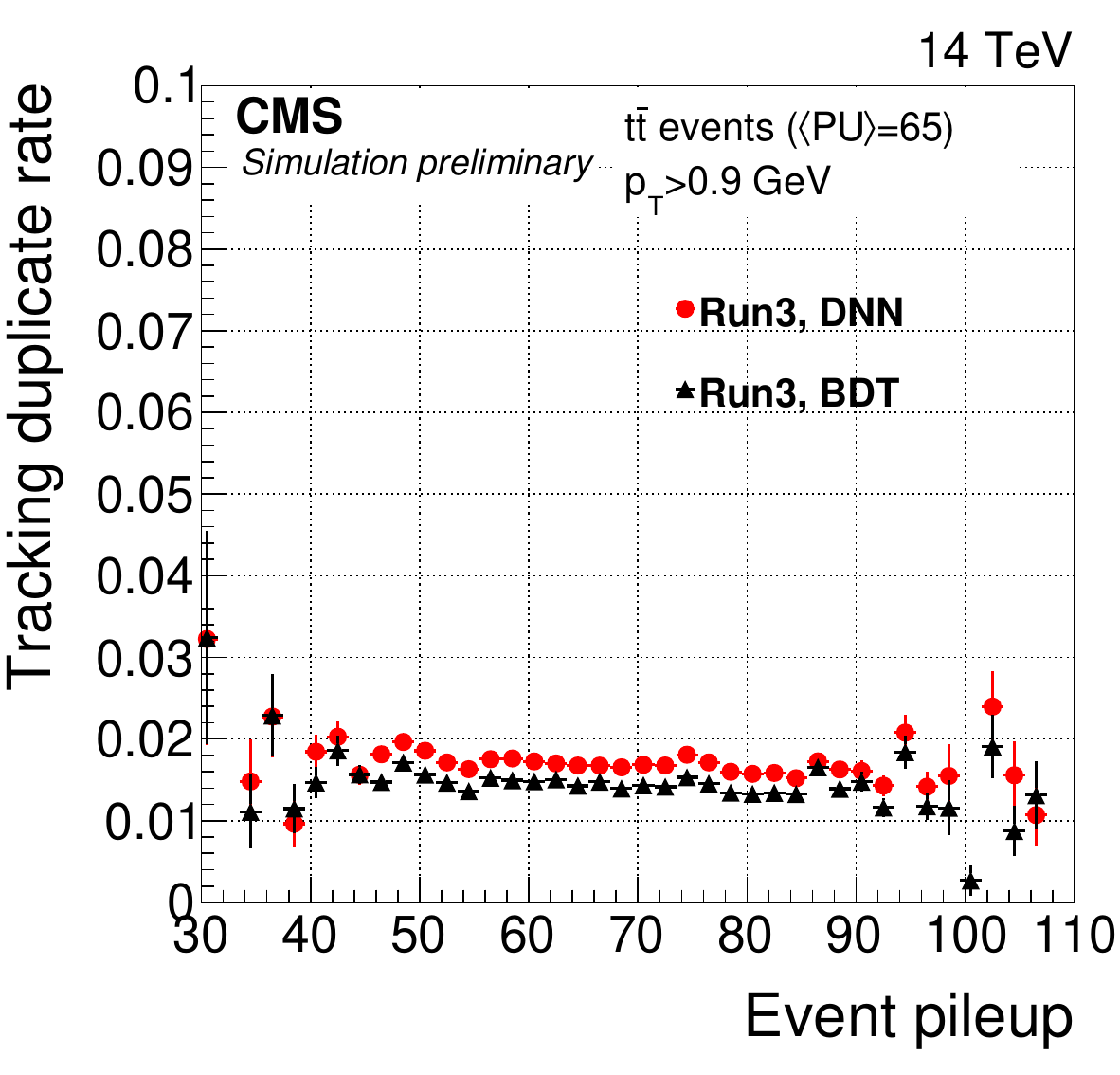}
    \caption{\label{fig:image3}}
    \end{subfigure}
\caption{(\subref{fig:image1}) The tracking efficiency is shown as a function of
the event pileup (or PU) for simulated tracks with p$_\text{T}$ > 0.9 GeV, $|\eta|$ < 3.0 and $|\text{d}_0|$ < 2.5 cm , (\subref{fig:image2}) tracking fake rate  as a function of the event pileup for tracks with p$_\text{T}>$0.9 GeV, (\subref{fig:image3}) tracking duplicate rate  as a function of the event pileup for tracks with p$_\text{T}$ > 0.9 GeV.}
\label{fig:pu}
\end{figure}

The efficiency, fake rate, duplicate rate versus pileup are shown in figure \ref{fig:pu}. The tracking efficiency when the DNN is used is consistent with or slightly higher than the one obtained using the BDT independently of the pileup. Overall the efficiency is increased by 1\%. The tracking fake rate when the DNN is used is lower than the one obtained using the BDT across the full pileup range, with a reduction up to about 30\% for higher pileup values. The tracking duplicate rate when the DNN is used is higher than the one obtained using the BDT. The increase of about 20\% is consistent across the entire pileup range.\\

\begin{figure}[h!]
\centering
    \begin{subfigure}[b]{0.3\textwidth}
    \centering
    \includegraphics[width=\textwidth]{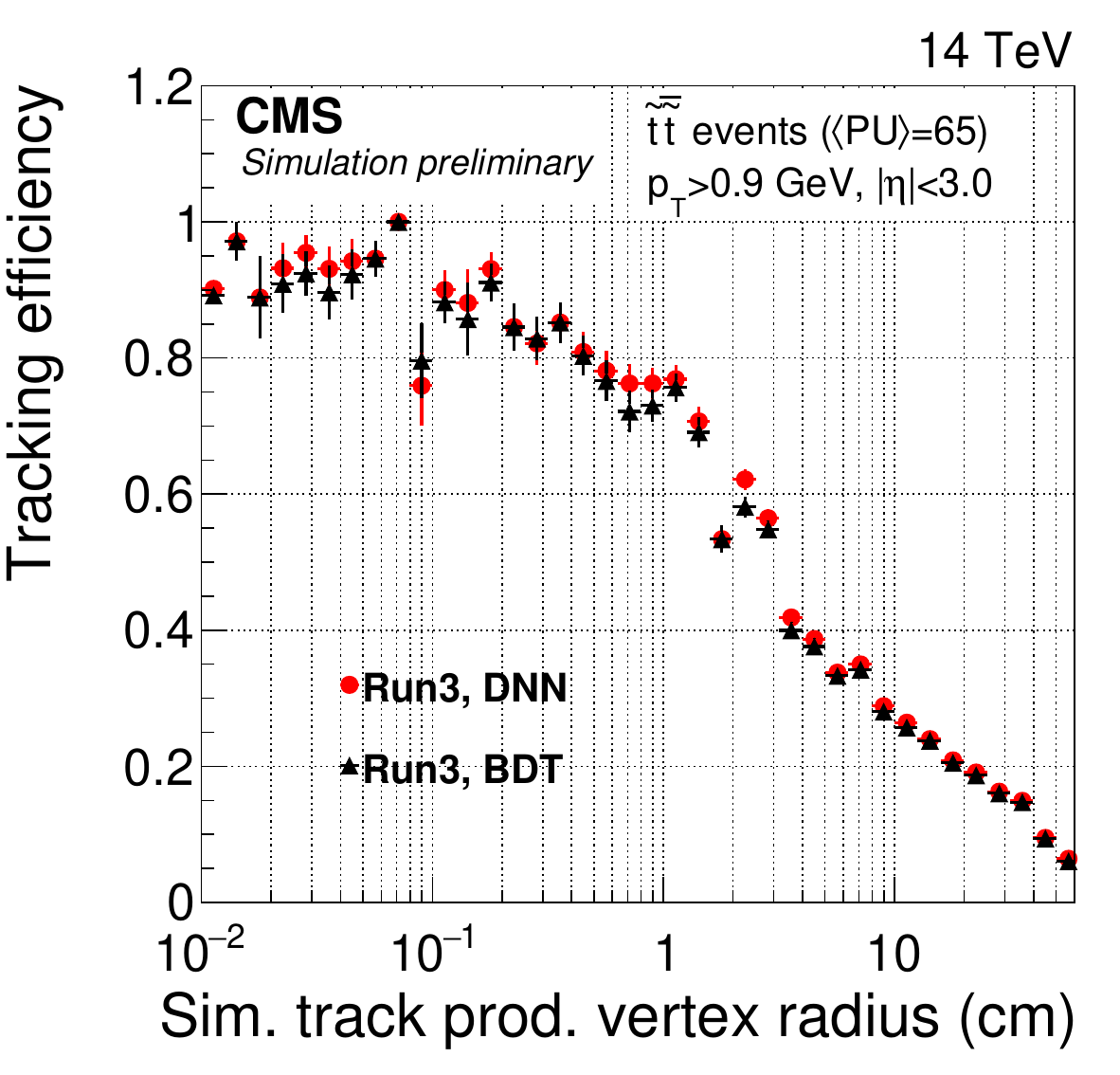}
    \caption{\label{fig:image1}}
    \end{subfigure}
\quad
    \begin{subfigure}[b]{0.3\textwidth}
    \centering
    \includegraphics[width=\textwidth]{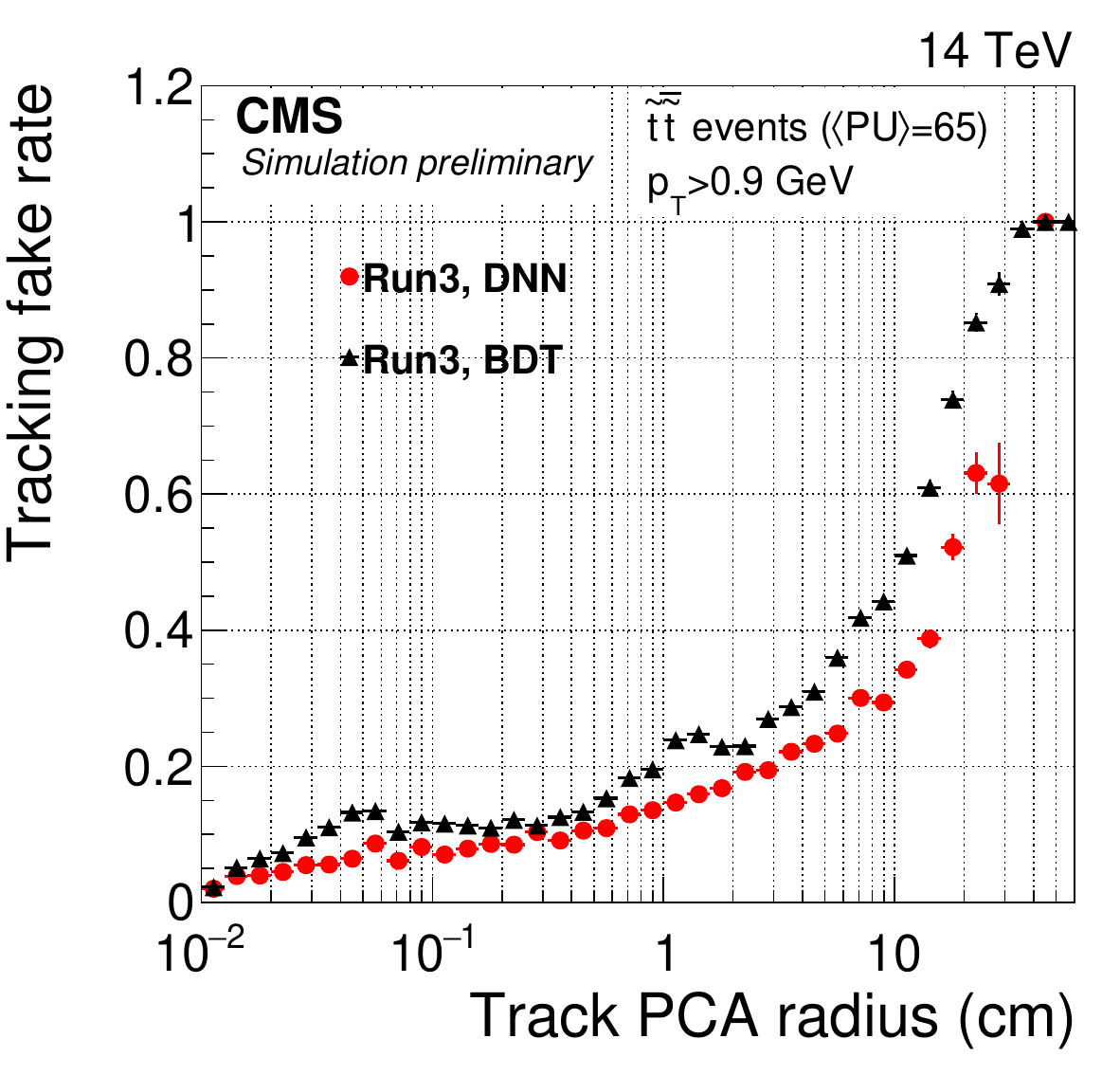}
    \caption{\label{fig:image2}}
    \end{subfigure}
\quad
    \begin{subfigure}[b]{0.3\textwidth}
    \centering
    \includegraphics[width=\textwidth]{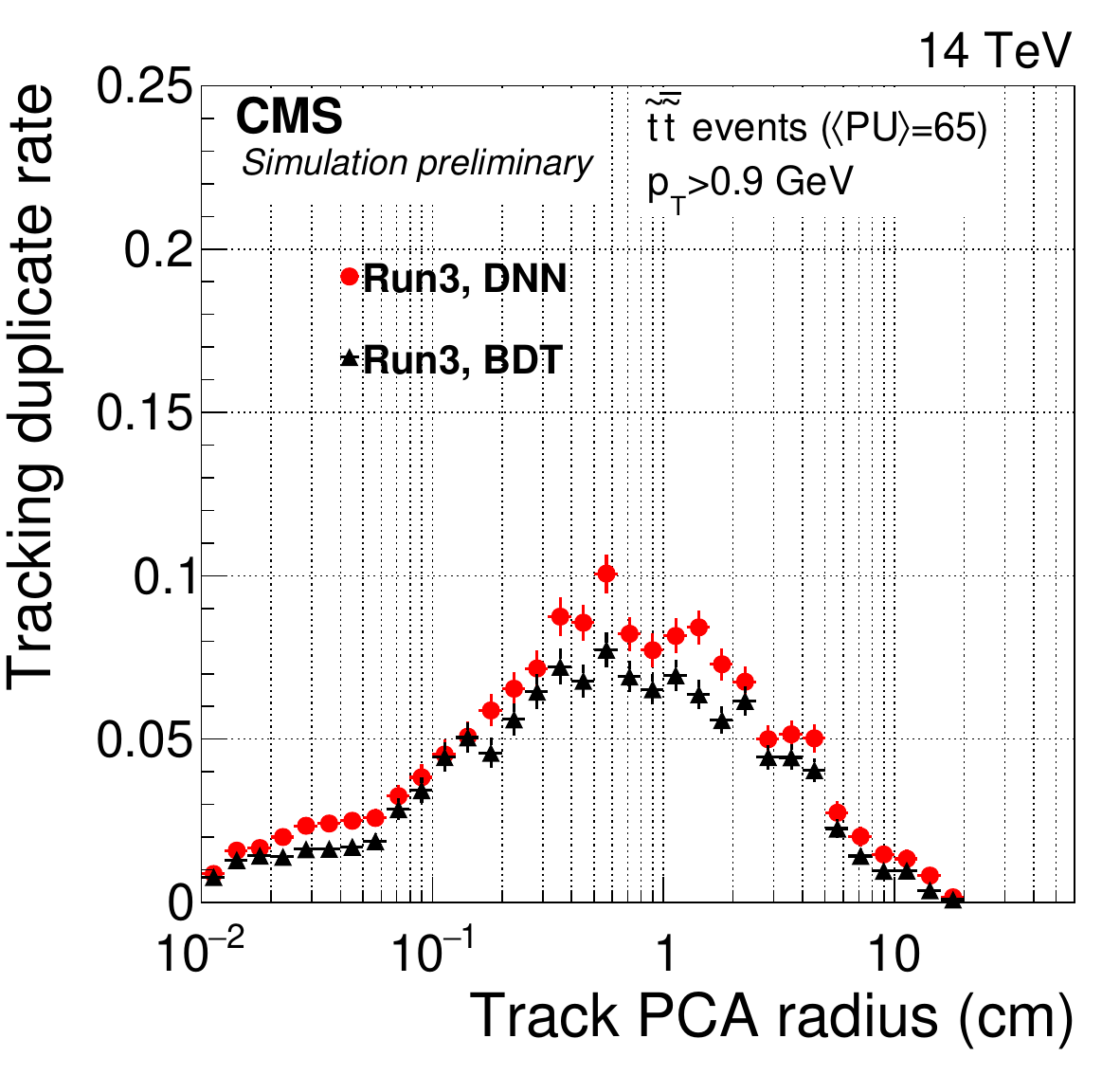}
    \caption{\label{fig:image3}}
    \end{subfigure}
\caption{ (\subref{fig:image1}) Tracking efficiency is shown as a function of
the simulated track production radius for simulated tracks with p$_\text{T}$ > 0.9 GeV, $|\eta|$ < 3.0, (\subref{fig:image2}) tracking fake rate as a function of
the radius of the track point of closest approach to the beamline (or d$_0$) for tracks with p$_\text{T}$ > 0.9 GeV, (\subref{fig:image3}) tracking duplicate rate as a function of the radius of the track point of d$_0$ for tracks with p$_\text{T}$ > 0.9 GeV.} \label{stop}
\end{figure}

The tracking efficiency when the DNN is used is consistent with or slightly higher than the one obtained using the BDT at all track origin radii. One can notice the
higher statistics for high radius values (> 1cm) in the $\tilde{\text{t}}\bar{\tilde{\text{t}}}$ sample, as evident by decreasing statistical uncertainties for the corresponding bins. The tracking fake rate when the DNN is used is lower than the one obtained using the BDT across all the radii values, with a reduction of about 30\%.
The tracking duplicate rate when the DNN is used is higher than the one obtained using the BDT for all the radii, by about 20\% (see figure \ref{stop}).

\section{Timing}

The CPU time of the DNN evaluation is measured as a fraction of the total tracking time. The DNN timing is compared to the timing of the Run 2 BDT. A slight speedup is observed with batch size 1, i.e. evaluating the DNN track by track as for the BDT, and 
a more pronounced speedup is observed with larger batches at evaluation time.
%a larger one with larger batches at evaluation time. 
A batch size of 16 is chosen due to memory footprint constraints and is the value currently used in the offline reconstruction. The results for the BDT and the DNN with batch sizes 1 and 16 are reported in table \ref{tab-1}.
\begin{table}[h!]
\centering
\caption{Evaluation of the track selection timing as a fraction of the total tracking time.}
\label{tab-1}       % Give a unique label
% For LaTeX tables you can use
\begin{tabular}{ll}
\hline
Method & \% of the total tracking time \\
\hline
BDT & 4.9 \\
DNN - batch size 1 & 3.4\\
DNN - batch size 16 & 0.9\\
\hline
\end{tabular}
% Or use
\vspace*{5cm}  % with the correct table height
\end{table}

\section{Summary}
The CMS Collaboration improved the track selection by means of a simple feed-forward DNN. The DNN leverages a larger training dataset and improves the efficiency and fake rate with respect to the previous BDT selection. The evaluation time is also faster, thus slightly reducing the total tracking CPU time.


\begin{thebibliography}{}

\bibitem{CMS:2014pgm}CMS Collaboration, Description and performance of track and primary-vertex reconstruction with the CMS tracker. {\em JINST}. \textbf{9}, P10009 (2014)

\bibitem{CMS-DP-2023-009}CMS Collaboration, Performance of the track selection DNN in Run 3. (http://cds.cern.ch/record/2854696, 2023)

\bibitem{CMS:2017yfk}CMS Collaboration, A. \& Others Particle-flow reconstruction and global event description with the CMS detector. {\em JINST}. \textbf{12}, P10003 (2017)

\bibitem{Lantz:2020yqe}Lantz, S. \& Others, Speeding up particle track reconstruction using a parallel Kalman filter algorithm. {\em JINST}. \textbf{15}, P09030 (2020)

\bibitem{CMS-DP-2022-018}CMS Collaboration, Performance of Run 3 track reconstruction with the mkFit algorithm. (http://cds.cern.ch/record/2814000
, 2022)

\bibitem{clevert2016fast}Clevert, D., Unterthiner, T. \& Hochreiter, S., Fast and Accurate Deep Network Learning by Exponential Linear Units (ELUs).  (2016)

\bibitem{chollet2015keras}Chollet, F. \& Others, Keras. (https://keras.io, 2015)

\bibitem{abadi2016tensorflow}Martín Abadi \& Others, TensorFlow: Large-Scale Machine Learning on Heterogeneous Distributed Systems. (2016)

\bibitem{kingma2017adam}Kingma, D. \& Ba, J., Adam: A Method for Stochastic Optimization.(2017)


\end{thebibliography}
\end{document}